# Morphological entropy encodes cellular migration strategies on multiple length scales

Yanping Liu[1,2], Yang Jiao[3,4], Qihui Fan[5], Xinwei Li[1,2], Zhichao Liu[1,2], Jun Hu[6], Jianwei Shuai[7,8,9,*], Liyu Liu[10,*], Zhangyong Li[1,2,*].

[1] Department of Biomedical Engineering and [2] Chongqing Key Laboratory of Big Data for Bio Intelligence, Chongqing University of Posts and Telecommunications, Chongqing 400065, China

[3] Materials Science and Engineering and [4] Department of Physics, Arizona State University, Tempe, Arizona

[5] Beijing National Laboratory for Condensed Matter Physics and CAS Key Laboratory of Soft Matter Physics, Institute of Physics, Chinese Academy of Sciences, Beijing 100190, China

[6] Department of Neurology, Southwest Hospital, Army Medical University, Chongqing 400038, China

[7] Department of Physics and [8] Fujian Provincial Key Laboratory for Soft Functional Materials Research, Xiamen University, Xiamen 361005, China

[9] Wenzhou Institute, University of Chinese Academy of Sciences, Wenzhou 325000, China

[10] Chongqing Key Laboratory of Soft Condensed Matter Physics and Smart Materials, College of Physics, Chongqing University, Chongqing 401331, China

*Corresponding authors: jianweishuai@xmu.edu.cn, lyliu@cqu.edu.cn, lizy@cqupt.edu.cn

**ABSTRACT**

Cell migration is crucial to many physiological and pathological processes. During migration, a cell adapts its morphology, including the overall morphology and nucleus morphology, in response to various cues in complex microenvironments, e.g. topotaxis and chemotaxis. Thus, cellular morphology dynamics can encode migration strategies based on which various migration mechanisms can be inferred. However, how to decipher cell migration mechanisms encoded in the morphology dynamics remains a challenging problem. Here we introduce a novel universal metric, namely cell morphological entropy (CME), by combining parametric morphological analysis with Shannon entropy. The utility of CME, which accurately quantifies the complex cellular morphology on multiple length scales through the deviation from the perfect circular shape, is demonstrated using a variety of normal and tumorous cell lines in distinct in vitro microenvironments. Our results reveal that 1) the effects of geometric constraints on cell nucleus, 2) the emerging interplays of MCF-10A cells migrating on collagen gel, and 3) the critical transition of tumor spheroid from proliferation to invasion. The analysis indicates that the CME offers a physically interpretable and efficient tool to quantify morphology on multiple length scales in real-time, which provides more insights into cell migration, and further contributing to the understanding of the diverse behavioral modes as well as collective cell motility in more complex microenvironment.

**Significance**: We introduce a novel approach based on cell morphological entropy to analyze morphological dynamics of cell migration regulated by various environmental cues, which enables us to accurately quantify the morphological dynamics and decipher the migration strategies encoded therein. We also demonstrate that the superior utility of the approach in revealing mechanisms underlying cell migration on multiple length scales.

**INTRODUCTION**

Cell migration plays an important role in the normal development of tissues or organs, including wound healing [1-3], immune response [4], and embryogenesis [5]. Also, many human diseases are mainly dominated by ill-regulated cell migration, such as cancer invasion and metastasis [6, 7].

Usually, cells migrating in complex microenvironment are regulated by environmental cues [8] and intracellular signaling pathways [9], and thus exhibiting diverse modes of cell migration [10, 11]. For example, chemotaxis mediated by diffusible cues [12], haptotaxis in response to surface-bound chemical cues [13], durotaxis in response to differences in substrate stiffness [14], and etc. More specifically, MDA-MB-231 cells in a two-state micropatterns exhibit a limit cycle, while MCF-10A cells show excitable bistable dynamics [15]. Interestingly, cells reversing, following and sliding past each other upon collision in this micropatterns, have been observed [16]. Similarly, oriented collagen fibers in microenvironment can stabilize cellular protrusions and further guide 3D cell migration [17]. During directed migration, the corresponding persistence is exponentially correlated with migration speed, this typical relationship is not only dominated by actin flows, but also regulated by Arp2/3 complex that supports lamellipodia extension [18-20]. Besides the migration dynamics, cells also exhibit distinct morphological dynamics when responding to different external cues. For example, cell nuclei are stretched for overcoming the steric hindrance caused by physical constraints [21], which strongly correlates with nuclear envelope stretch-sensitive proteins [22]. Since both cell morphology and modes of cell migration are the consequences resulted from a combination of extracellular cues and intracellular signals, they are closely related to each other. For instance, the elongated mode of migration termed as 'mesenchymal', is dominated by the actin polymerization that pushes the plasma membrane forward, while the rounded mode termed as 'amoeboid' is mainly dependent on actomyosin contractility [23-25]. Therefore, how to in turn reveal the characteristics of cell migration and/or environmental cues from the perspective of cell morphological dynamics is a challenging problem.

In order to decipher mechanisms underlying cell migration based on the most vivid cell morphology, many novel researches have been carried out. For example, cellular morphology neural networks are constructed to identify subcellular compartments and the cell types of neuron reconstructions [26]. Also, machine learning is employed to classify cell shape into different phenotypes, demonstrating that morphological phenotype controlled by ECM mechanics and Rho/ROCK-signaling, facilitates cancer cells to navigate non-uniform ECM [27]. Recent work shows that morphological classes of single cell–derived clones obtained from 216 features of cell and nucleus using unsupervised clustering analysis, predict distinct tumorigenic and metastatic potentials in vivo [28]. Moreover, shape fluctuations of chromatin globule surface and nuclear envelope are both thermally and actively driven, and decreasing amplitudes can serves as a reliable cell cycle stage indicator [29]. In our previous work, a quantitative approach is developed to describe major characteristics of morphology of tumor-cell spheroid, and verifies the ability of DDR1 inhibitor 7rh to weaken the invasion of single cells [30]. Taken together, there are increasing evidences in studying cell morphology and its relationship with cell functions and migration modes, and enabling to obtain more insights into the mechanisms underlying cell migration. However, the approaches utilized mainly focus on a specific feature and pre-determined combination of features of the cell morphology, which may be not sufficient to explore the emerging properties from the ensemble distributions of those features. Additionally, these approaches are mainly developed based on machine learning algorithm and lots of morphological data, which may be limited in analyzing time-varying morphological features. Therefore, how to quantify morphological features of cell or nucleus in real-time using a simple and unified approach becomes a big challenge.

Here we introduce a theoretical metric based on a combination of morphological analysis and Shannon entropy, to describe the morphological dynamics of cells that are mediated by complex physical or biochemical cues. To this end, we obtained and analyzed different types of cell experimental data. We find that for different length scales of cell data including cell nucleus, single cell and cell spheroid, the approach can measure accurately the changes in morphology, especially the angular and radial features. By analyzing the time-dependent CME components, some information or mechanisms encoded in the changes in morphology could be captured, e.g., the dynamics of cell nucleus when overcoming the steric hindrance of ECM, the interplays of MCF-10A cells migrating on the 3D collagen gel, and the transition of tumor spheroid from proliferation to invasion under the regulation of DDR1 inhibitor 7rh. Thus, the CME metric enables us to explore the mechanisms underlying cell migration in pathophysiological environments, such as cancer and other physiological conditions.

**MATERIALS AND METHODS**

**Morphological analysis**

For better understanding of the process of developing CME approach, we first process the time-lapsed images and extract the boundary of 2D morphology of research object (see Fig. S1 for more details in Supplementary Material). Secondly, mapping the morphology in Cartesian coordinate system (CCS) into Polar coordinates system (PCS), and move the centroid of morphology to the origin of PCS. Then, we obtain the coordinates $(r_i, \theta_i)$ of the boundary, which accurately describe the shape of research object. Here the subscript $i$ denotes the numerical order of each point at the boundary. After obtaining the coordinates $(r_i, \theta_i)$, it is necessary to set a number lag $dN$ and then the coordinates $(r_{i+dN}, \theta_{i+dN})$ are selected to form a new boundary $\mathbf{M} = (r_j, \theta_j)$, see Fig. S3 in Supplementary Material for the detailed discussion of how to determine the value of $dN$. Undoubtedly, the new boundary discards some fluctuations caused mainly by imaging noises, but still represents the main shape of research object. Next, the displacements between any two consecutive points are computed by $\Delta \mathbf{M} = \mathbf{M}_j - \mathbf{M}_{j-1}$, which can also be presented by two terms, i.e., radial $\Delta r_j$ and angular $\Delta \theta_j$ components. Subsequently, the probability density functions (PDFs) of the two components are derived from their statistical histograms, i.e., $p(\Delta r)$ and $p(\Delta \theta)$, as seen in Fig. 1A. The detailed description can be found in the latter section.

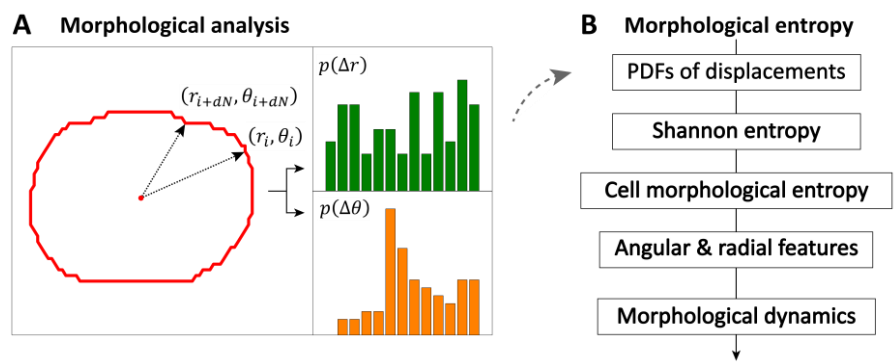

FIGURE 1 Flowchart of constructing cell morphological entropy. (A) Morphological analysis of the research object in polar coordinates. The green and orange bars present probability density functions (PDFs) of radial and angular displacements, respectively. (B) Derivation of cell morphological entropy based on PDFs and Shannon entropy.

**Cell morphological entropy**

Inspired by the relationship above, we further introduce Shannon entropy to develop a robust description method for quantifying the changes in morphology, which we termed as 'cellular morphology entropy' (see flowchart in Fig. 1B). Entropy is an extensively used concept in thermodynamics, and typically used to describe the degree of disorder or randomness in the states of molecules. It was not until 1948 that it was introduced to describe the 'uncertainty' in information source by C. E. Shannon [31]. Thus, it is also referred to as 'Shannon entropy' when involves in information theory. For a random event with probabilities of occurrence $p_1, p_2, ..., p_n$, the corresponding Shannon entropy can be derived from the following formula:

$$H = -\sum_{i=1}^{n} p_i log2(p_i),$$

where $n$ is the total number of events that possibly occur, $p_i$ is the probability of occurrence for each event, and $log2()$ is the logarithm function with a base of 2. The Shannon entropy not only measures how much 'choice' is involved in the selection of event, but also indicates how 'uncertain' the outcome of event could be. According to the characteristics of logarithm function above, it is not difficult to deduce that entropic value will be maximal when probability $p_i$ is identical to each other, i.e., $p_i = 1/n$, meaning that one cannot judge which event is most likely to occur [32]. For comparing the results for different cases, it is essential to normalize all entropic values by dividing the $H_{max}$, as a consequence, all entropic values will be rescaled into a close interval of [0,1]. Next, we further substitute the $p_i$ with $p(\Delta\theta)$ or $p(\Delta r)$ separately, and obtain the $H$ for each component. According to spirits of Shannon entropy and PDFs of the components, it is obvious that the more regular (or irregular) the shape of object, the closer the corresponding $H$ is to 0 (or 1). For avoiding confusion caused by abbreviations, we use 'CME' to denote Shannon entropy $H$ that correlates with the morphology of object.

**Biophysical interpretations of CME**

In order to definitely illustrate the biophysical interpretations of CME approach, we analyzed the morphology of two types of single-cell migration following the procedure stated in Fig. 1, i.e., amoeboid and mesenchymal modes of migration (see inset in Fig. 2A), and the corresponding results are exhibited in Fig. 2. Evidently, the PDFs of angular displacement for the two types of morphologies possess the different trends as a whole, i.e., the probability values for amoeboid mode locate in the small interval of 0.0 ~ 0.05 while the values for mesenchymal mode cover a large range of -0.05 ~ 0.15 (Fig. 2A). Consequently, the difference leads to a narrower PDF of angular displacement for amoeboid mode in contrast to that for mesenchymal mode, which indicates theoretically that the blebs (or protrusions) of amoeboid mode distribute on the angular direction in a more uniform manner. It's noting that the term 'narrower' is used when the PDF is far away from a uniform distribution. Similarly, the PDF of radial displacement for amoeboid mode is narrower when compared with that for mesenchymal mode (Fig. 2B), thus also showing that the blebs of amoeboid mode distribute on the radial direction in a more uniform manner. In addition, the CME components (i.e., CMEa for angular and CMEr for radial features) of the two types of morphologies also exhibit significant differences, namely, the CMEa for amoeboid mode is smaller markedly than that for mesenchymal mode, and the CMEr for the former is also less slightly than that for the latter but with statistical significance ($*** p < 0.001$) (Fig. 2C). Here the values of CME components are mainly determined by the natural features of cell morphology, thus it is highly possible that one could utilize this metric to distinguish the modes of cell migration. For deeply exploring the performance of CME approach, we additionally analyzed multiple and single lamellipodia with similar features, which differs from the significant differences between amoeboid and mesenchymal modes. The results agree well with our judgment and further verify the effectiveness of CME in capturing some subtle differences of morphology. See Fig. S2 in Supplementary Material for more detailed discussions.

Taken together, the CME can be used to measure angular and radial features of a given morphology, and the more uniformly distributed the features, the smaller the value of CME. For better understanding the relationship mentioned above, we conclude the following two aspects: 1) when there are multiple unobvious features (e.g., blebs) on the angular direction, we could consider they distribute more uniformly on this direction and result in a narrower PDF (smaller CMEa); 2) when there are significant differences in the features (e.g., protrusions) on the radial direction, we also consider they distribute less uniformly on this direction and result in a wider PDF (larger CMEr). In other words, the CMEa and CMEr describe the heterogeneity of angular and radial features of cell morphology, respectively.

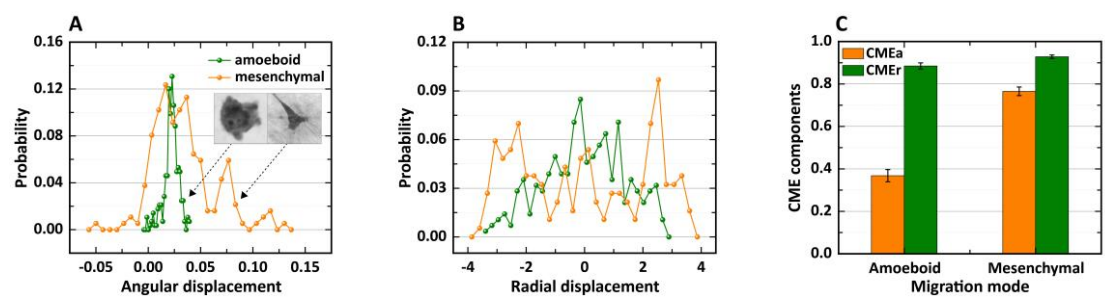

FIGURE 2 Biophysical interpretations of CME. (A) PDFs of angular displacement of cell morphology. The inset shows representative single-cell migration modes adapted from the work [33], with permission. (B) PDFs of radial displacement of cell morphology. (C) CME components of the two types of migration modes. Data are presented as mean ± sd.; n=10, 8 for amoeboid and mesenchymal modes, respectively; $***\, p < 0.001$; $t$-test. It shows that the difference between CMEa (or CMEr) for the two modes is statistically significant (not shown).

**In vitro cell experiments**

*Microstructural array experiment.* A microstructural array consisting of sequentially channels and chambers was designed and fabricated, and the width of channel decreases gradually from 11.2 to 1.7 μm. To better observe and analyze the dynamical process of cells squeezing through the array, cell nuclei are stained with 1.5 μg/mL Hoechst (in red) and imaged in a sampling time of 0.2 min. It should be noted that the images of cell nuclei analyzed in this study are obtained from the attached videos of the work [21].

*3D collagen gel experiment.* In vitro cell migration experiment was carried out as follows: MCF-10A cells first were obtained from China Infrastructure of Cell Line Resource (Beijing, China) and marked with green fluorescent protein (GFP). Additionally, the culture medium used here is Dulbecco's modified Eagle's medium-F12 (Corning, Corning, NY), which was also supplemented with 1% penicillin/streptomycin (Corning), 5% horse serum (Gibco, Gaithersburg, MD), 20 ng/mL human EGF (Gibco), 100 ng/mL cholera toxin (Sigma-Aldrich, St. Louis, MO), 0.5 mg/mL hydrocortisone (Sigma-Aldrich), and 10 mg/mL insulin (Roche Diagnostics, Basel, Switzerland). Next, Type I collagen extracted from rat tail tendon (Corning) was diluted and neutralized to pH ~7.1, then the collagen solution was spread on the substrate of petri dish and incubated in 37°C for 30 min until it polymerized into a 3D matrix with a concentration of 2 mg/mL and thickness of ~2 mm. Subsequently, 0.5 mL of cell suspension was dropped on top of collagen gel and formed a randomly distributed cell population with a low cell density of $10^4$ cells/$cm^2$ after incubated for 2 h. Finally, both a confocal laser scanning microscope and an automatic inverted fluorescent microscope (Nikon Ti-E, Tokyo, Japan) were applied to obtain the time-lapse images of cells with a sampling time of 2 min. See works [32, 34] for more details of the experiment.

*Cell spheroid experiment.* The experiment was performed based on three types of tumor cells, i.e., U87 (glioma tumor) cells, H1299 (lung cancer) cells and MDA-MB-231 (invasive breast cancer) cells, and all of them first were marked by green fluorescent protein (GFP). Then, cell suspension with a density of $1.0×10^4$ cells/mL was prepared and further seeded into an ultra-low attachment (ULA) plate with 96-well round bottom where cell spheroid formed after 96 h. Next, the spheroid was transferred to a new 96-well flat bottom ULA plate containing culture medium with a collagen concentration of 2 mg/mL, where the spheroid was imaged by a CCD camera (Neo 5.5 sCMOS, Andor, USD) for further analysis. See the work [30] for more details of the experiment.

**Statistical analysis**

Statistical analysis is performed using custom MATLAB (R2018b, USA). When data satisfy two criteria of normality and equal variance, parametric tests are used: *t*-test for two groups and ANOVA (analysis of variance) for more than two. If data processed by bijective transformation, still do not satisfy the criteria, non-parametric tests are applied: Wilcoxon rank sum for two groups and Kruskal–Wallis for more than two. Differences are significant at confidence levels greater than 95% (two-tailed). Three levels of significance are distinguished, i.e., $* p < 0.05$; $** p < 0.01$; $*** p < 0.001$, determined by the standard Michelin Guide scale.

Correlation coefficient reflects the degree of correlation between two variables. When data are continuous numerical variables and both meet normality (or possess obvious single-peaks), the Pearson coefficient is preferred, and if the data do not satisfy the normality after transformation, Spearman or Kendall coefficient are optional. See more details in the work [32].

**RESULTS**

**Morphological dynamics of cell nucleus squeezing through micro-structured channel array**

In the previous subsection, we have clearly clarified the biophysical interpretations of CME on the basis of the cell morphologies of amoeboid and mesenchymal modes. Now, we further apply the CME approach to investigate changes in morphology of nucleus when cell squeezes through spatial constraints. Here the images of cell nucleus are obtained from the work by Fabry et al. [21], in which the authors studied 3D migration in confined microenvironment with different stiffness (see more interesting results in the works [21, 35]).

*Features of channel array characterized by CME.* In this study, we first transmit the video of cell migration into a series of images, and analyze the time-lapse images by utilizing CME approach, and the corresponding results are exhibited in Fig. 3. It clearly indicates that one cell nucleus is squeezing through a narrow channel (as marked by yellow arrows) at different times, and the nucleus is 'rod-like' because of the physical constraints with a gradually decreasing width from 8.4 to 6.6 μm (Fig. 3A). Furthermore, the migration speed seems like stable qualitatively because the nucleus goes through the chamber at the time interval of ~ 50 min, as denoted by the labels in vertical axis of Fig. 3A. In terms of quantitative analysis, both of the CME components (CMEa and CMEr) possess the same characteristics of 'peak and valley' (Spearman's coefficient = 0.77), i.e., four peaks and three valleys, exhibiting visibly the effects of these channels and chambers on cell nucleus, respectively (Fig. 3B). In addition, CME components also behave differently, that is, the values of peak and valley for CMEa increase gradually, as indicated by the dotted line. However, the values of peak for CMEr almost keep stable around 0.58, which also differs greatly from the increase in those of valley. The changing trends above of CME components further illustrate that CMEa and CMEr response differently when encountering the same external cues from ECM, which may be related to the nuclear envelope [22]. In order to assess the changes in cell nucleus as a whole, we also average the CME components and obtain the resulting average CME following the similar trend with that of CMEa (Fig. 3C).

Even if there are distinct differences between CME components, CMEa still correlates strongly with CMEr (Pearson's coefficient = 0.90), for instance, CMEr increases gradually with the increasing of CMEa, and the relationship between them could be fitted well by a linear variation

'y=0.50x+0.25' ($R^2$=0.80), as indicated in Fig. 3D. According to the features of distribution, we further divide the scatters in Fig. 3D into three clusters (marked by I, II and III) with K-means clustering algorithm [36], and the averaged values for each cluster are plotted in Fig. 3E, among which the error bars denote sd. (standard deviation) for the two CME components. It is highly evident that CMEa significantly increases from 0.52 to 0.62 and 0.68, while CMEr first increases from 0.51 to 0.58 and then keeps stable around 0.58, thus forming three clusters for CMEa and two clusters for CMEr. On the one hand, the three clusters of CMEa in Fig. 3E could be explained perfectly by three migration states in microstructured array: 1) cluster I demonstrates that the cell nucleus is locating in chamber and possess smaller CMEa as it has more space to recover from the highly confined state; 2) cluster II shows that the cell nucleus is entering (or exiting) the channel and the confinement is increasing (or decreasing) gradually; 3) cluster III illustrates that the cell nucleus is squeezing through the channel and possess greater CMEa because of the stronger physical confinement. On the other hand, the two clusters of CMEr could be utilized to identify different structures, i.e., the large CMEr corresponds to channel while the small to the chamber. In addition, the two aspects above also directly mirror that CMEa is more sensitive to space confinement when compared with CMEr.

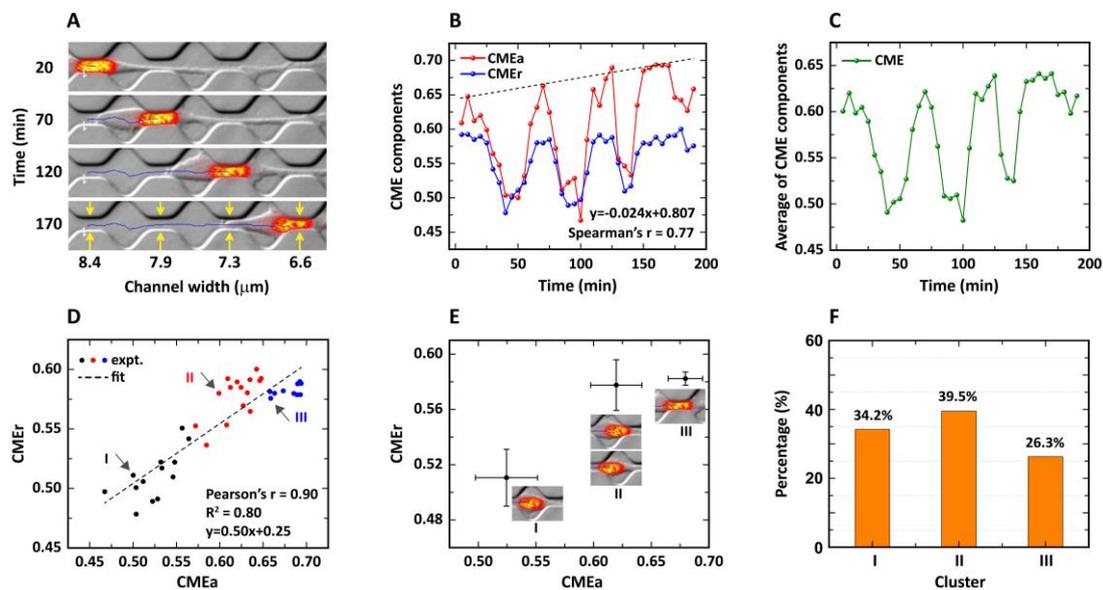

FIGURE 3 Morphological dynamics of cell nucleus in micro-structured channel array. (A) Breast cancer cells migrating in microstructure consisting of sequentially channels and chambers adapted from the work [21] with permission. Nuclei are stained with Hoechst and shown in red. (B) CME components of angular (red) and radial (blue) displacements as functions of time. The dotted line is auxiliary to eyes. (C) Average of CME components in (B) as a function of time. (D) Scatters of CMEr vs. CMEa. The dotted line is a linear fit to experimental data, which are divided into three clusters (see black, red and blue points) with K-means clustering. (E) Three states indicated by scatters of CMEr vs. CMEa. Data are presented as mean ± sd. (F) Percentage of scatters for each cluster (I~III).

***Estimation of steric hindrance of channel array.*** Finally, we count the number of scatters in each cluster and plot the histogram in Fig. 3F. The results show that the percentages for cluster I~III are 34.2%, 39.5% and 26.3%, respectively, which are extremely consistent with the results (34.2%, 44.7% and 21.1%), determined by artificial judgment. Due to the unchanged sampling time $\Delta t$ = 5 min, the percentage here could also be regarded as (or equivalent to) dwell-time of cell nucleus in a special structure. According to the array with three chambers and four channels traveled by cell nucleus, we roughly assess the time when nucleus locates in chambers and channels and their ratio is about 0.75 (3/4), which should be larger for the fact that the horizontal length (23 μm) of chamber is greater than that (18 μm) of channel, theoretically. Actually, the time ratio of cluster (I)/(III) equals to 1.3, which is 73.3% larger than 0.75. When considering the contribution of cluster II based on cluster III, the ratio of cluster (I)/(II+III) 0.52 is significantly 31.6% less than the theoretical value. Thus, it is reasonable to deduce two key aspects: 1) the partly deformed nucleus corresponding to cluster II does cost more time in contrast to the completely deformed state for cluster III (also see Figs. S4, S5); 2) the channels exert an influence (at least 31.6%) on impeding cell migration when compared with that of chambers. The results show that geometrical confinements in ECM can deform cell nucleus and further hinder cell migration to some extent.

Additionally, we also investigated morphological dynamics of other MDA-MB-231 cells in more narrower channels, especially the widths of 4.4 μm and 1.7 μm (Figs. S4 and S5). The cell nuclei exhibit similar changes in morphology with that shown in Fig. 3A when squeezing through the highly constrained channels, and possess a larger CMEa, i.e., 0.6949 for 4.4 μm and 0.7496 for 1.7 μm (see orange bars in Fig. 4B). Except for the similarities above, there are some subtle differences, for example, the cell nucleus first remains the shape close to circle and continuously quivers before entering the channel with a width of 1.7 μm, then elongates along the channel just like a chopstick when locating at the channel, finally the elongated morphology returns to circle-like status after exiting the channel. See Supplementary Material for more detailed analysis.

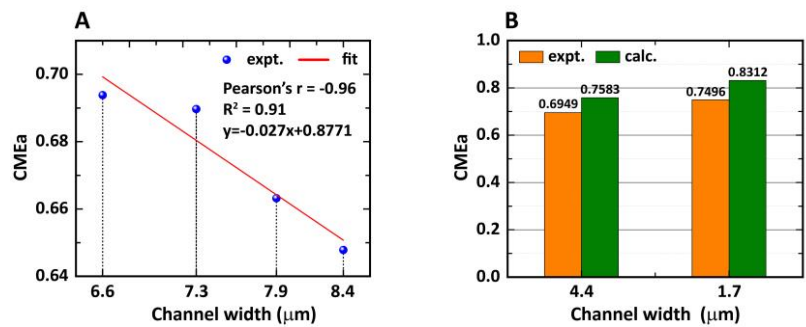

FIGURE 4 Quantitative relationship between CMEa and the width of channel. (A) Maximal CMEa of cell nucleus as a function of the width (8.4 ~ 6.6 μm) of channels. The red line is a linear fit to experimental CMEa (blue points), denoted by 'y=-0.027x+0.8771' ($R^2$=0.91). (B) Maximal CMEa of cell nucleus in more narrower channels with widths of 4.4 and 1.7 μm. The orange bars denote CMEa obtained from experimental nucleus of MDA-MB-231 cell, while the green bars present results calculated from formula in (A).

***Correlation between CMEa and the width of channel.*** To quantitatively illustrate the effects of geometrical confinements on the morphology of cell nucleus, we further analyzed the relationship between the widths of channels and the CMEa of nucleus migrating in those channels. It is noted that the CMEa (not CMEr) is chosen because of its larger sensitivity (see Fig. 3E). The channels here are structures shown in Fig. 3A, and the CMEa are maximal values at t=10, 70, 125 and 160 min in Fig. 3B. On the basis of the characteristics of maximal values, we introduced plainly a linear function 'y=kx+b' (x and y correspond to width and CMEa, respectively) to fit these values, and finally derived the formula 'y=-0.027x+0.8771' ($R^2$=0.91) (Fig. 4A). To further validate the universality of relationship described by the formula above, we subsequently used it to calculate theoretical CMEa, attaining 0.7583 and 0.8312 for channels with widths of 4.4 μm and 1.7 μm, respectively (see green bars in Fig. 4B). It is extremely obvious that both the theoretical values are larger (9.12% and 10.9%) than those obtained from experiments, which we believe may be explained by the following aspects: 1) the morphology of cell nucleus is correlated with the sizes of geometrical confinements, non-linearly; 2) although all of cell nuclei tested correspond to MDA-MB231 breast carcinoma cell line, they perhaps behave differently in response to the same external cues because of individual differences; 3) the data analyzed is not sufficient because there is a limit when requesting data from other works. The first and second hypotheses above are reasonable due to the complexities of living matter or biological systems to some extent, which may provide some guidance for further verification in the future.

**Emerging interplays of MCF-10A cells migrating on 3D thick collagen gel**

*Alternate changes in morphology of a pair of cells.* In order to validate the utility and efficiency of CME approach, we continue to investigate the morphology of MCF-10A cell migrating on 3D thick collagen gel. Fig. 5A shows representative time-lapsed images of a pair of cells labeled by 'up' and 'down', obviously indicating the two cells migrate toward to each other (as indicated by yellow arrows). After applying the CME approach to cellular morphology, we obtained CME components of the cell pair as functions of time (Figs. 5B-D). In Fig. 5B, the CME components of the up cell possess the almost same trends (Spearman's coefficient = 0.91), i.e., the values first increase and then decrease, and reach 'stable' maximum (~0.7 for CMEa, ~0.6 for CMEr) in the time interval of about 40~70 min. However, the CME components of the down cell exhibit significant differences from those of the up cell (Fig. 5C), namely, first fluctuate at a high level of about 0.6~0.7 for CMEa (or about 0.4~0.5 for CMEr), then start to decrease steeply from about 30 min and reach the minimum at about 50 min (~0.4 for CMEa, ~0.25 for CMEr), subsequently increase in reverse until about 70 min and finally return to the high level before decreasing. In this process, CMEa changes synchronously with CMEr, which is also quantified by Spearman's coefficient = 0.85. In addition, the values of CMEr are generally less than those of CMEa for the two cells, indicating that angular characteristic encoded by CMEa is more sensitive than radial characteristic by CMEr when cell responses to the same external cues from microenvironment (see the similar results in Fig. 3).

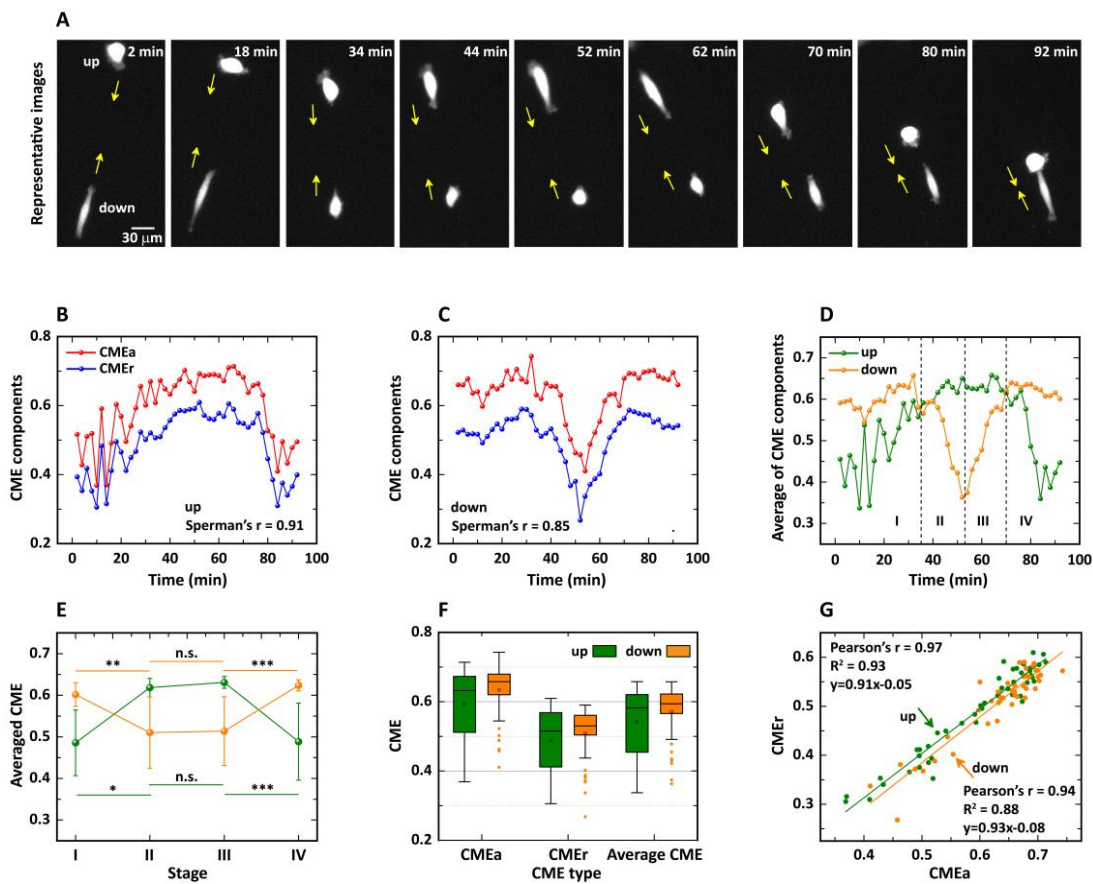

FIGURE 5 Emerging interplay of MCF-10A cells migrating on thick 3D collagen gel. (A) Brightfield image series of a typical pair of cells migrating on collagen gel, adapted from the work [34] with permission. (B) CME components of the up cell as functions of time. The red and blue lines represent CMEa and CMEr, respectively, and the Spearman's coefficient is 0.91. (C) CME components of the down cell as functions of time. The Spearman's coefficient is 0.85. (D) Average of CME components of the up cell (green line) and of the down cell (orange line). The average CME are divided artificially into four stages according to the tendencies, i.e, stage I (2~34 min), II (36~52 min), III (54~70 min) and IV (70~92 min). (E) Averaged CME for each stage in (D). Data are presented as mean ± sd.; $**\ p < 0.01$, $***\ p < 0.001$, n.s. means 'not significant'; Kruskal-Wallis test. (F) Boxplot of CME components and their average CME. The box plot indicates the mean (small square in the box), the median (black line in the box), 25th percentile (bottom line of the box), 75th percentile (top line of the box), and 1.5*IQR (Interquartile range, bars). (G) Scatters of CMEr vs. CMEa. The Pearson's coefficients are 0.97 for the up and 0.94 for the down cells. The green and orange lines are linear fits to the corresponding scatters, obeying functions 'y=0.91x-0.05' ($R^2$=0.93) and 'y=0.93x-0.08' ($R^2$=0.88), respectively.

***Symmetry and similarities in alternate changes.*** Further, the average of CMEa and CMEr vividly presents the differences between the up and down cells, as seen in Fig. 5D. In order to better compare the behavioral modes of the two cells during different migration periods, we divide the average CME into four stages (as marked by vertical dotted lines) based on its characteristics with time lapsing. In stage I, i.e., 2~34 min (or stage IV, 70~92 min), the CME of the down cell are slightly increased (or decreased) and are greater than the gradually increased (or decreased) CME of the up cell, whereas the sharply changed former are significantly less than the basically stable latter in stage II, 36~52 min (or stage III, 54~70 min). Next, all values in each stage are averaged and plotted in Fig. 5E to quantitatively validate the descriptions above. The results clearly show two kinds of symmetry: 1) the changing trend of averaged CME of the down cell is almost opposed to that of the up cell; 2) all values in stages I and II seem like symmetry with those in stages III and IV.

Besides the time-varying features of CME, the box plot in Fig. 5F also shows that not only all the averages of CMEa, CMEr and average CME of the down cell are a bit greater than those of the up cell, but also the CME of the up cell are more sparsely distributed than those of the down cell, illustrating directly the statistical differences in morphology between the two cells. Even so, there are still some interesting similarities, such as the linear variations of CMEr vs. CMEa, following formulas 'y=0.91x+0.05' ($R^2$=0.93) and 'y=0.93x+0.09' ($R^2$=0.88) pretty well, respectively for the up and down cells.

***A potential indicator of cell forces?*** On the basis of results above, we suggested that the behavioral modes encoded in morphology may embody interplays between a pair of cells, especially the force exerted by single cells. In previous works [32, 34, 37], we observed that the active tensile forces generated by migrating cells can remodel collagen fibers, which is directly verified by the phenomenon that elongated cells contribute to the formation of fiber bundles while rounded cells doesn't re-organize the surrounding collagen, and in turn, the fiber bundles bridging two cells typically regulate cell migration and lead to strongly correlated motility. Therefore, the CME could be viewed as a potential indicator that is used to further measure, how cells exert force to the surrounding environments, accurately and real-time. See more detailed analysis in the Discussion section.

In addition to the behavioral mode of the two migrating cells descripted above, we also observed another two phenomena described by CME (Figs. S6, S7), i.e., 1) two cells that are close to each other change their morphologies synchronously and possess roughly the same morphological characteristics; 2) one cell keeps its morphology unchanged while another cell changes the morphology dramatically, finally also approaching to each other. See Supplementary Material for more detailed analysis.

## Critical transition of tumor spheroid from proliferation to invasion

***Transition detected by CME from proliferation to invasion.*** Except for the applications of CME metric in analyzing morphologies of cell nucleus and cell pairs, we also explored the proliferation and invasion of three types of cell spheroids, namely H1299 (lung cancer), MDA-MB-231 (breast cancer) and U87 (glioma tumor), using CME approach that differs from the complex methods (e.g., perimeter[2]/area ratio) in our previous work [30]. Fig. 6A shows representative images of U87 cell spheroid without 7rh (DDR1 inhibitor) treatment, vividly exhibiting the changes in morphology of spheroid, for instance, some 'fingers' formed by single cells occur at the boundary as time goes by, as indicated by the white arrow (see Fig. S8 for more details). Subsequently, we calculated the average of CMEa and CMEr for H1299 cell spheroid (n=3 independent experiments), as shown in Fig. 6B, where the CME for no-7rh and with-7rh cases possess the similar trends, that is, first remain stable with small fluctuations and then increase gradually. The stable stage manifests that although cells start to proliferate and lead to an expansion of the morphology of original spheroid, which does not affect the shape until the presence of 'fingers'. Here the fingers denote the invasion of cancer cells away from the spheroid, which may be driven by the hypoxic and acidic tumor microenvironment [38]. In addition, the transition from 'stable' to 'increase' stage (i.e., from proliferation to invasion) for no-7rh case is earlier than that for with-7rh case, meaning that DDR1 inhibitor 7rh could effectively inhibit the transition. To further validate the results, we continue to analyze experimental data for MDA-MB-231 and U87 cell spheroids (n=3 independent experiments for each type of cell spheroid). Evidently, all transitions for no-7rh cases are earlier, indicating that 7rh does inhibit the transition from proliferation to invasion, regardless of cell types (Figs. 6C-D).

It should be noted that all CME of with-7rh case are greater than those of no-7rh case for H1299 cell spheroid, however the relation 'greater' becomes 'smaller' and 'roughly equal' for MDA-MB-231 and U87 cell spheroids, which may be caused by different sizes (average radius) of original cell spheroids (t=0 min). Here, we suggest the differences in CME relations for different cases produce less effects on the transition results above, because they are mainly determined by the slopes (or inflection points) of CME profiles, rather than the magnitudes.

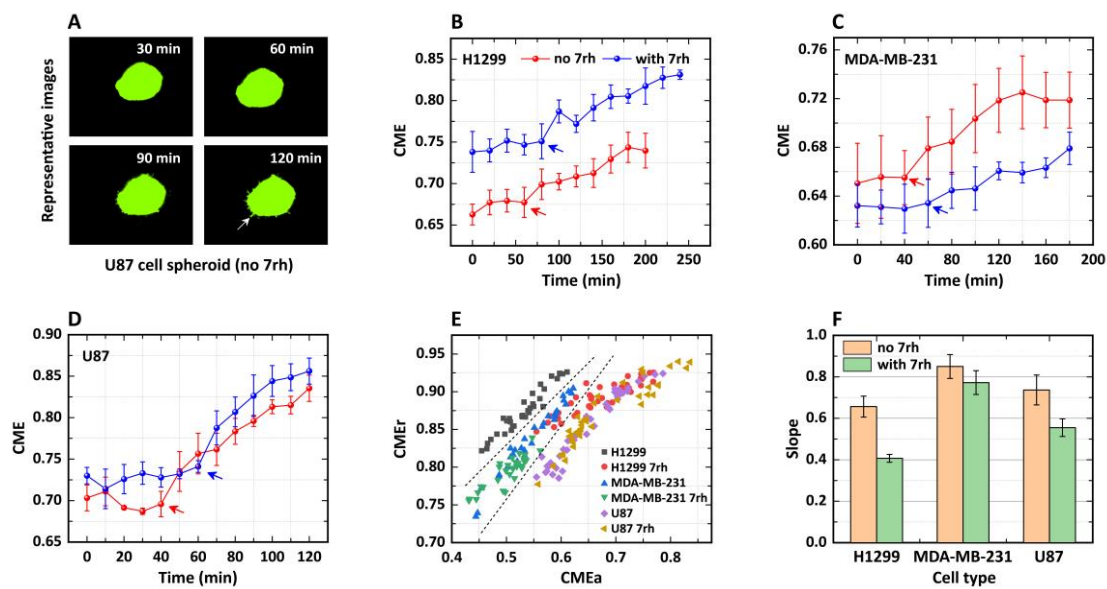

FIGURE 6 Transitions detected by CME from proliferation to invasion of cell spheroids. (A) Representative fluorescent images of U87 cell spheroid without 7rh treatment. (B) CME of H1299 cell as a function of time. The red and blue lines are corresponding to cases without 7rh and with 7rh treatments. (C, D) CME of MDA-MB-231 and U87 cells as functions of time. (E) Scatters of CMEr

vs. CMEa for three types of tumor spheroids. (F) Slopes of linear fits to six cases in (E). Data are presented as mean ± s.e.m. (standard error of the sample mean), n=3 independent experiments.

***Inhibited transition derived from CME scatters.*** Different from the results shown by Fig. 5G, the scatter of CMEr vs. CMEa indicates that significant differences are observed in different cell types (Fig. 6E). For no-7rh cases, the scatters for H1299 and U87 distribute relatively at the upper-left and bottom-right areas, respectively, while those for MDA-MB-231 locate at the region sandwiched by the scatters for other two cell types, as marked by dotted lines. Additionally, for a given CMEa (for example CMEa=0.6), H1299 has the largest CMEr, followed by MDA-MB-231 and U87, which indicates that the heterogeneity in length of fingers is the most obvious in H1299 cell spheroids. When treated by 7rh, the CMEr and CMEa are affected significantly, especially H1299 and MDA-MB-231 cells. For instance, 1) the scatters of with-7rh case for H1299 deviate greatly from those of no-7rh case, thus leading to two separate regions (see the dotted lines); 2) the scatters of with-7rh case for MDA-MB-231 possess smaller CMEr and CMEa, and forming a region with an area approximately half of the area of no-7rh case. To further explore the relationship between CMEr and CMEa, all the scatters for the three types of cell spheroids have been fitted by linear variations with the slopes plotted in Fig. 6F. The histograms vividly present the slopes of no-7rh cases are significantly larger than those of with-7rh cases, for H1299 and U87, while there is less difference for MDA-MB-231 after considering the s.e.m. errors (also see Fig. S9). The results above illustrate that 1) 7rh treatment can change the quantitative correlation of CMEr with CMEa, and inhibit the invasion of individual cells away from cell spheroids; 2) different types of cells have distinct sensitivities to 7rh, and resulting in the changes in slopes.

**DISCUSSION**

In this study, we have introduced an approach termed as cell morphological entropy, derived from a combination of morphological analysis and Shannon entropy, which enables us to analyze the angular and radial characteristics concerning the morphology relating to cells and further explore the mechanisms of cell migration underlying morphological dynamics.

We first investigate sub-cellular object, cell nucleus, that squeezes through a micro-structured array consisting of sequentially channels and chambers, among which the channels become narrower gradually. We found that the CME of nucleus in channels are significantly larger than those in chambers, and the changing trends of CME vividly reflect two characteristics of the array, i.e., channels connecting to chambers sequentially and the gradually narrowing channels. Therefore, it's probably feasible to view the CME approach as a powerful metric that is utilized directly to measure the characteristics of physical cues geometrically constraining adhesion sites [39], such as the oriented 3D matrix or 1D lines, when direct measurements are ineffective or appropriate tools are lacking. Moreover, the linear variation derived from a fit to the scatters of CMEa vs. width describes quantitatively relationship between the angular features of morphology for a given nucleus and the sizes of channels, which not only manifests the straightforward effects of physical constraints on the morphological dynamics of nucleus, but also characterizes the deformation ability or sensitivity to physical cues, to some extent. Further, these properties may be combined to construct an identification code, e.g., a fingerprint, for signifying the essential features of cell nucleus.

There are differences between theoretical values and calculated values in Fig. 4, when the linear variation for one nucleus is applied to analyze other nuclei. It may be explained by non-linear variation or individual differences, i.e., following a non-linear relation for many cells or a cell aggregate. Thus, it is necessary to analyze more cell nuclei in the array for obtaining systematic variations, and estimate the potential of these cells to overcome the similar barriers, further contributing to remodeling or engineering the microenvironment to hinder the invasion and metastasis of cancer cells.

Furthermore, we also divide the scatters of CMEr vs. CMEa into three clusters, which correspond to three stages, i.e., migration in channels, in chambers and entering/exiting the channels, respectively. Although the clustering is determined by the morphology of nucleus, it mirrors features of invasiveness/migration modes of cancer cells to some extent. For example, the time costed by nucleus are roughly assessed when performing these special migration modes, clearly indicating one key point: the time when entering the channels is greater than that when exiting the channels, illustrating the cell needs more time to coordinate the intracellular signaling pathways associated with nuclear envelope stretch-sensitive proteins for better adapting physical constraints. The analysis above agrees well with the mechanisms underlying cell responses to spatial constraints reported in previous works [22].

Except for the sub-cellular nucleus, we further study the morphology of MCF-10A cells migrating on the thick collagen gel and found the changes in morphology of two cells that are close to each other, exhibits obvious regularity, i.e., when the morphology of one cell is close to rounded status, another cell elongates deviating from the rounded status, vice versa. From another point of view, the changes possess the high degree of symmetry both in the vertical (magnitude) and horizontal (time) axis, as indicated by four stages in Fig. 5D, further illustrating that an alternant mode emerges from the morphologies of the two cells and the interplay mode may be mediated by a medium of communication, such as collagen fibers [34, 40] or biochemical factors [41]. Here, the principle of communication is similar with that used in communication of follower cells with leader cells through adhesion-based mechanical interactions [42]. When the changed morphology is related to contributions made by cells to meeting each other, it may refer to forces exerted on ECM [43, 44], energy costed [45] or other physical quantities. Interestingly, we also observed another two phenomena occurring during two cells approach to each other, i.e., 1) two cells change their morphology synchronously from elongated to rounded status and 2) one cell always remains elongated status while another cell keeps the shape closing to rounded status, which we term as 'synchronous' and 'unilateral' modes, respectively (see Figs. S6, S7). The two interplay modes are complementary to the alternant mode and beneficial to obtain more insights into multicellular motion, especially collective cell migration, but the modes need to be further studied and confirmed because of the insufficient data tested in Supplementary Material.

Due to the result reported in our previous work [34], that cellular morphology is strongly correlated with the force exerted on ECM, thus one could assess the characteristics of the force by utilizing CME approach when cell migrates in complex environments, including temporal and spatial aspects. In terms of the alternant mode in Fig. 5, we argue that the down cell first exerts a greater pulling force on collagen fibers, and the up cell gradually increases its force after sensing the tensile force when they're far apart (from 200 to 150 μm). As the distance between them decreases (from 150 to 100 μm), the force exerted by the up cell starts to increases and exceeds the force by the down cell. Finally, the force exerted by the down cell becomes the main contribution again when they are closer (from 100 to 50 μm). The alternant change may enable one cell to sense/judge the status of another cell and further adapt its migration mode, contributing to improving the efficiency of communication or correlated movement. In addition, the description method above could be used to analyze another two modes in Supplementary Material. We believe that the emerging interplays encoded in morphology embody the diversities of cell-cell communications to some extent, which may contribute to explaining some collective cell migration [46], such as wound healing, histogenesis and the invasion and metastasis of cancer cell.

Finally, we analyze the morphology of three types of tumor cell spheroids for detecting the transitions from proliferation to invasion, and the results firstly illustrate that DDR1 inhibitor 7rh can change the quantitative correlation of CMEr with CMEa, and inhibit the invasion of individual cells away from cell spheroid, which may help us to obtain a better understanding of malignant mammary tumors that reorient the collagen fibers to be perpendicular to the breast gland and apply these structures as 'highways' for migration away from the crowded/dense regions occupied by epithelial cells [47, 48]. Secondly, different types of tumor cells have distinct sensitivities to 7rh and result in the changes in slopes of CME profiles, thus it's essential to optimize the dosage of 7rh to control the invasion of tumor cells, avoiding or lessening the development of drug resistance. In addition, 7rh here can be replaced by other biochemical factors, such as epithelial growth factor (EGF), batimastat and glucose, to explore the individual or superposed effects of these factors on cancer. Here, the CME approach combining with cell spheroid model may provide us a new platform for the screening and evaluation of effective candidate drugs in the era of personalized cancer therapy.

## CONCLUSIONS

In this article, we have proposed an approach termed as cell morphological entropy, which enables us to explore the mechanisms of cell migration underlying morphological dynamics. Our results have also validated the utility and efficiency of CME approach, which can be used to measure accurately the effects of geometric constraints on cell nucleus that is viewed prevailingly as the main source of steric hindrance for 3D invasion, especially the positive correlation of the CME of nucleus with the size of constraint. Furthermore, the interplays of MCF-10A cells migrating on thick collagen gel also emerge from the changes in morphology characterized by CME, which not only illustrates the diversities of cell-cell interactions, but also emphasizes the crucial role of collagen fibers in regulating cell migration behaviors. Finally, we also captured the transitions of three types of tumor cell spheroids from proliferation to invasion, and further confirmed the ability of DDR1 inhibitor 7rh to weaken the invasiveness of cancer cells. Overall, our approach contributes to analyzing the information encoded in morphology, and revealing cellular migration strategies on multiple length scales in complex microenvironment.

## AUTHOR CONTRIBUTIONS

## ACKNOWLEDGMENTS

We thank David B. Brückner at Institute of Science and Technology Austria for useful comments. This work is supported by the Natural Science Foundation of Chongqing, China [Grant No. CSTB2022NSCQ-MSX1260]; the National Natural Science Foundation of China [Grant Nos. 62106032, 62171073]; and the Key Project of Technology Innovation and Application Development of Chongqing, China [Grant No. cstc2021jscx-gksbX0060].

## DECLARATION OF INTEREST

The authors declare no competing interests.